\documentclass[article,12pt]{elsarticle}

\usepackage{epsfig}
\usepackage{amssymb}
\usepackage{amsmath}
\usepackage{rotating}
\usepackage{array}
\usepackage{tikz}
\newcommand{\LL}{$\Lambda$}
\newcommand{\LLs}{$\Lambda$ }
\newcommand{\BL}{$B_{\Lambda}$}
\newcommand{\BLs}{$B_{\Lambda}$ }

\journal{Nuclear Physics A}

\begin{document}

\begin{frontmatter}

\title{On the Binding Energy and the Charge Symmetry Breaking in $A \leq$ 16 \LL-hypernuclei}

\author[a,b]{E.~Botta}
\author[b]{T.~Bressani \corref{cor1}}
\author[b]{A.~Feliciello}

\address[a]{Dipartimento di Fisica, Universit\`a di Torino, via P. Giuria 1, Torino, Italy}
\address[b]{INFN Sezione di Torino, via P. Giuria 1, Torino, Italy}

\cortext[cor1]{Corresponding author. E-mail address: tullio.bressani@to.infn.it}

\begin{abstract}
In recent years, several experiments using magnetic spectrometers provided high precision results in the field of Hypernuclear Physics.
In particular, the accurate determination of the \LL-binding energy, \BL, contributed to stimulate considerably the discussion about the Charge Symmetry Breaking effect in \LL-hypernuclei isomultiplets.
\par
We have reorganized the results from the FINUDA experiment and we have obtained a series of \BLs values for \LL-hypernuclei with $A \leq$ 16 by taking into account data only from magnetic spectrometers implementing an absolute calibration of the energy scale (FINUDA at DA$\Phi$NE and electroproduction experiments at JLab and at MaMi).
We have then critically revisited the results obtained at KEK by the SKS Collaboration in order to make possible a direct comparison between data from experiments with and without such an absolute energy scale.
A synopsis of recent spectrometric measurements of \BLs is presented, including also emulsion experiment results.
\par
Several interesting conclusions are drawn, among which the equality within the errors of \BLs for the $A$ = 7, 12, 16 isomultiplets, based only on recent spectrometric data.
This observation is in nice agreement with a recent theoretical prediction.
\par
Ideas for possible new measurements which should improve the present experimental knowledge are finally put forward.

\end{abstract}


\end{frontmatter}


\section{Introduction}
\label{intro}

The binding energy \BLs of a \LLs hyperon in a \LL-hypernucleus (shortened as hypernucleus in the following), $^{A}_{\Lambda}$Z, is the most  straightforward observable which characterizes such a strange nuclear system.
It is defined as:

\begin{equation}
 B_{\Lambda} =  [M(\Lambda) + M(^{A-1}Z) - M(^{A}_{\Lambda}Z)]\ c^{2}
\label{BL}
\end{equation}
where $M(\Lambda)$ is the mass of the \LLs hyperon, $M(^{A-1}Z)$ is the mass of the core nucleus in its ground state and $M(^{A}_{\Lambda}Z)$ is the mass of the hypernucleus.
$M(^{A}_{\Lambda}Z)$ is determined by means of the measurement of the missing mass in production reactions with magnetic spectrometers or by the sum of the masses and of the kinetic energies of the decay products in measurements with photographic emulsions or bubble chambers.
Table~\ref{tab:tab1} contains the most complete series of experimental determinations of \BLs in the case of $s$- and $p$-shell hypernuclei.
It is the starting point of the discussion presented in this paper.

\begin{sidewaystable}[p]
\caption{Synopsis of the experimental values of \BLs for A $\leq$ 16 hypernuclei. Column 1: hypernucleus; column 2: emulsions;  column 3: KEK-SKS; column 4: revised KEK-SKS; column 5: DA$\Phi$NE-FINUDA; column 6: electroproduction. References are in parentheses; [t.w.] stands for this work. In columns 2--6 the first error is statistical, the second one is systematic; in columns 5 and 6 the error quoted for results from Ref.~\protect{\cite{ref:plb698}} and, respectively, Ref.~\protect{\cite{ref:cusanno}} is total.}
\vspace{3mm}
\hspace{-10mm}
\label{tab:tab1}
\setlength{\extrarowheight}{1.0pt}
\begin{small}
\begin{tabular}{|cccccc|}
\hline
 & emulsions (MeV) & $(\pi^{+}, K^{+})$ (MeV) & $(\pi^{+}, K^{+})$ (MeV) & $(K^{-}_{stop}, \pi^{-})$ (MeV) & $(e, e^{\prime} K^{+})$ (MeV) \\
 & & KEK-SKS~\cite{ref:hashi} & KEK-SKS revised [t.w.] & DA$\Phi$NE-FINUDA & JLab, MaMi \\
\hline \hline
$^{3}_{\Lambda}$H   & 0.13$\pm$0.05$\pm$0.04~\cite{ref:juric,ref:davis} & & & & \\
\hline
$^{4}_{\Lambda}$H   & 2.04$\pm$0.04$\pm$0.04~\cite{ref:juric,ref:davis} & & & & 2.157$\pm$0.005$\pm$0.077~\cite{ref:esser} \\
\hline
$^{4}_{\Lambda}$He  & 2.39$\pm$0.03$\pm$0.04~\cite{ref:juric,ref:davis} & & & & \\
\hline
$^{5}_{\Lambda}$He  & 3.12$\pm$0.02$\pm$0.04~\cite{ref:juric,ref:davis} & & & & \\
\hline
$^{6}_{\Lambda}$H   & & & & 4.0$\pm$1.1~\cite{ref:npa881,ref:prl108} & \\
\hline
$^{6}_{\Lambda}$He  & 4.25$\pm$0.10~\cite{ref:juric} & & & & \\
                    & 4.18$\pm$0.10$\pm$0.04~\cite{ref:davis} & & & & \\
\hline
$^{7}_{\Lambda}$He  & & & & & 5.55$\pm$0.10$\pm$0.11~\cite{ref:gogami1} \\
\hline
$^{7}_{\Lambda}$Li  & 5.58$\pm$0.03$\pm$0.04~\cite{ref:juric,ref:davis} & 5.22$\pm$0.08$\pm$0.36 & 5.82$\pm$0.08$\pm$0.08 &
                    5.85$\pm$0.13$\pm$0.10~\cite{ref:plb681},[t.w.] & \\
                    & & & & 5.8$\pm$0.4~\cite{ref:plb698} & \\
\hline
$^{7}_{\Lambda}$Li$^*$ & 5.26$\pm$0.03$\pm$0.04 & 4.90$\pm$0.08$\pm$0.36 & 5.50$\pm$0.08$\pm$0.08 & 5.53$\pm$0.13$\pm$0.10 & \\
~\cite{ref:tamura}                       & & & & 5.48$\pm$0.40 & \\
\hline
$^{7}_{\Lambda}$Be  & 5.16$\pm$0.08$\pm$0.04~\cite{ref:juric,ref:davis} & & & & \\
\hline
$^{8}_{\Lambda}$He  & 7.16$\pm$0.70$\pm$0.04~\cite{ref:juric,ref:davis} & & & & \\
\hline
$^{8}_{\Lambda}$Li  & 6.80$\pm$0.03$\pm$0.04~\cite{ref:juric,ref:davis} & & & & \\
\hline
$^{8}_{\Lambda}$Be  & 6.84$\pm$0.05$\pm$0.04~\cite{ref:juric,ref:davis} & & & & \\
\hline
$^{9}_{\Lambda}$Li  & 8.53$\pm$0.15~\cite{ref:juric} & & & & 8.36$\pm$0.08$\pm$0.08~\cite{ref:urciuoli} \\
                    & 8.51$\pm$0.12$\pm$0.04~\cite{ref:davis} & & & & \\
\hline
$^{9}_{\Lambda}$Be  & 6.71$\pm$0.04$\pm$0.04~\cite{ref:juric,ref:davis} & 5.99$\pm$0.07$\pm$0.36 & 6.59$\pm$0.07$\pm$0.08 & 6.30$\pm$0.10$\pm$0.10~\cite{ref:plb681},[t.w.] & \\
                    & & & & 6.2$\pm$0.4~\cite{ref:plb698} & \\
\hline
$^{9}_{\Lambda}$B   & 7.88$\pm$0.15~\cite{ref:juric} & & & & \\
                    & 8.29$\pm$0.18$\pm$0.04~\cite{ref:davis} & & & & \\
\hline
$^{10}_{\ \Lambda}$Be & 9.30$\pm$0.26~\cite{ref:juric} & & & & 8.60$\pm$0.07$\pm$0.16~\cite{ref:gogami2} \\
                    & 9.11$\pm$0.22$\pm$0.04~\cite{ref:davis} & & & & \\
\hline
$^{10}_{\ \Lambda}$B  & 8.89$\pm$0.12$\pm$0.04~\cite{ref:juric,ref:davis} & 8.1$\pm$0.1$\pm$0.5 & 8.7$\pm$0.1$\pm$0.08 & & \\
\hline
$^{11}_{\ \Lambda}$B  & 10.24$\pm$0.05$\pm$0.04~\cite{ref:juric,ref:davis} & & & 10.28$\pm$0.2$\pm$0.4 [t.w.] & \\
\hline
$^{12}_{\ \Lambda}$B  & 11.37$\pm$0.06$\pm$0.04~\cite{ref:juric,ref:davis} & & & & 11.524$\pm$0.019$\pm$0.013~\cite{ref:tang} \\
\hline
$^{12}_{\ \Lambda}$C  & 10.76$\pm$0.19$\pm$0.04~\cite{ref:davis} & 10.80 fixed & & 11.57$\pm$0.04$\pm$0.10~\cite{ref:plb681},[t.w.] & \\
                    & & & & 10.94$\pm$0.06$\pm$0.50~\cite{ref:plb622} & \\
\hline
$^{13}_{\ \Lambda}$C  & 11.22$\pm$0.08~\cite{ref:juric} & 11.38$\pm$0.05$\pm$0.36 & 11.98$\pm$0.05$\pm$0.08 & 11.0$\pm$0.4~\cite{ref:plb698} & \\
                    & 11.69$\pm$0.12$\pm$0.04~\cite{ref:davis} & & & & \\
\hline
$^{14}_{\ \Lambda}$C  & 12.17$\pm$0.33$\pm$0.04~\cite{ref:davis} & & & & \\
\hline
$^{15}_{\ \Lambda}$N  & 13.59$\pm$0.15$\pm$0.04~\cite{ref:juric,ref:davis} & & & 13.8$\pm$0.7$\pm$1.0 [t.w.] & \\
\hline
$^{16}_{\ \Lambda}$N  & & & & & 13.76$\pm$0.16~\cite{ref:cusanno} \\
\hline
$^{16}_{\ \Lambda}$O  & & 12.42$\pm$0.05$\pm$0.36 & 13.02$\pm$0.05$\pm$0.08 & 13.4$\pm$0.4~\cite{ref:plb698} & \\

\hline
\end{tabular}
\end{small}
\end{sidewaystable}
\par
At the dawn of Hypernuclear Physics \BLs was measured by analyzing the events produced in stacks of photographic emulsions by the interaction of $K^{-}$'s, both stopped and in flight.
The emulsion technique demonstrated remarkable performance in recognizing all the charged products of the disintegration of a hypernucleus and in measuring their energies.
By taking advantage of these capabilities, it was possible to determine \BLs accurately for hypernuclei in the 3 $\leq A \leq$ 15 range.
Ref.~\cite{ref:juric} summarizes the results achieved up to 1972.
They are listed in column 2 of Table~\ref{tab:tab1}; the quoted errors, of the order of percent, are only statistical.
In a successive compilation some data were confirmed, some others were updated and two new entries ($^{12}_{\ \Lambda}$C and $^{14}_{\ \Lambda}$C) were added, as the result of experimental efforts after 1972~\cite{ref:davis}.
In addition, a systematic error of $\pm$40 keV was assumed for the \BLs of each hypernucleus included in the compilation.
In Sec.~\ref{sec3} we will comment on this point more extensively.
The results from the compilation of Ref.~\cite{ref:davis} are listed as well in column 2 of Table~{\ref{tab:tab1}.
 In this case we quote both statistical and systematic errors.
\par
We have added to the list the T = 1 excited state of $_\Lambda^7$Li, labeled $_\Lambda^7$Li$^*$, relevant for the discussion about the T = 1, $A$ = 7 hypernuclear isotriplet (ground state of $_\Lambda^7$He, $_\Lambda^7$Be and $_\Lambda^7$Li$^*$) which will be developed in Sec.~\ref{sec4}.
\BL ($^{7}_{\Lambda}$Li$^{*}$) has been calculated, following Ref.~\cite{ref:nakamura}, by means of the energy spacing information from the $\gamma$-ray measurement, Ex(T = 1, 1/2$^{+}$) = 3.88 MeV~\cite{ref:tamura}, and of the excitation energy of $^{6}$Li$^{*}$(T = 1) = 3.56 MeV.
Consequently, 0.32 MeV have been subtracted from the \BLs values reported in Table~\ref{tab:tab1} for the ground state of $^{7}_{\Lambda}$Li.
The results for $_\Lambda^7$Li$^*$ are reported on a separate line.
\par
For more than forty years the investigation of the characteristic features of hypernuclei, \BLs in particular, has been carried out at different Laboratories by using magnetic spectrometers optimized for the study of the two-body reactions $(K^{-}, \pi^{-})$ and $(\pi^{+}, K^{+})$ on
nuclear targets $^{A}$Z, leading to the production of the corresponding hypernuclei $^{A}_{\Lambda}Z$.
A series of recent review papers~\cite{ref:hashi,ref:epja,ref:feli,ref:gal1} provides a good account of the experimental techniques and of the results obtained so far.

The Superconducting Kaon Spectrometer (SKS) Collaboration at the KEK-PS provided the largest amount of data. In particular, the \BLs values for eleven hypernuclei spanning over the 7 $ \leq A \leq $ 208 range were measured, the attention being focused on the light system sector $(A \leq 16)$.
In column 3 of Table~\ref{tab:tab1} we report the measured \BLs of these light hypernuclei, with the corresponding statistical and systematic
errors~\cite{ref:hashi}. It appears that the statistical errors are of the order of percent, while the systematic ones are at least about five times larger and reach up to 500 keV. The energy resolution ranged between 1.9 and 2.3 MeV FWHM~\cite{ref:hase}. We remind that it was not possible to use an elementary reaction for the calibration of the \BLs scale since no free neutron target is available. The energy calibration was obtained by adjusting $B_\Lambda$ of the $^{12}_{\ \Lambda}$C ground state to the value provided by previous emulsion measurements~\cite{ref:davis}. Being the reference, the excitation spectrum of $^{12}_{\ \Lambda}$C was then measured with an improved resolution of 1.45 MeV FWHM (thanks to the use of a thin target)~\cite{ref:hotchi}. The bulk of data from SKS on the spectroscopy of hypernuclei has been an invaluable input for theoretical studies on the \LL$N$ potential and on other topics of Hypernuclear Physics.

In the last few years, high quality data were produced at JLab on hypernuclei $^{A}_{\Lambda}$(Z-1), formed through the electroproduction reaction $(e, e^{\prime} K^{+})$ on nuclear targets $^{A}$Z. They are the neutron-rich, isotopic mirrors of those obtained with the aforesaid meson-induced two-body reactions.
The resolution achieved on the excitation energy spectra (ranging from 0.54 to 0.8 MeV FWHM) was more than a factor of two better than those obtained in the case of the meson-induced production reactions. The absolute energy scale was calibrated by studying the electroproduction reaction of \LLs and $\Sigma^{0}$ on a free proton target. The \BLs values for $^{7}_{\Lambda}$He, $^{9}_{\Lambda}$Li, $^{10}_{\ \Lambda}$Be, $^{12}_{\ \Lambda}$B and $^{16}_{\ \Lambda}$N are given in Refs.~\cite{ref:gogami1,ref:urciuoli,ref:gogami2,ref:tang,ref:cusanno}, respectively, and are listed in column 6 of Table~\ref{tab:tab1}.
Statistical and systematic errors are reported separately, when available. Also the recent result on $B_{\Lambda}(^{4}_{\Lambda} \mathrm{H})$, obtained by exploiting for the first time the high resolution $\pi$ decay spectroscopy at MaMi~\cite{ref:esser}, has been included.

The precise data coming from electroproduction experiments made possible the comparison of the \BLs values for isomultiplets with different values of $A$.
By defining
\begin{equation}
\Delta B_{\Lambda}(A,Z) = B_{\Lambda}(^{A}_{\Lambda}Z) - B_{\Lambda}(^{A}_{\Lambda}(Z-1))
\label{DBL}
\end{equation}
and by considering the values from SKS and JLab reported in Table~\ref{tab:tab1}, it appears that $\Delta B_{\Lambda}(12,6)$ amounts to $\sim$ $-$700 keV and
$\Delta B_{\Lambda}(16,8)$ to $\sim$ $-$1300 keV. The $\Delta B_{\Lambda}(A,Z)$ values are related to the Charge Symmetry Breaking (CSB) in the \LL$N$ interaction and suggest an unexpectedly large violation, very hard to be explained theoretically.
As recently shown by Gal~\cite{ref:gal2}, negative $\Delta B_{\Lambda}$ values of the order of 100--200 keV are expected for $p$-shell hypernuclei, with opposite sign and absolute value lower than $\Delta B_{\Lambda}(4,2)$ ($\sim$ +350 keV). Ref.~\cite{ref:gal2} contains as well an updated list of theoretical contributions by other authors which we won't repeat here.
A similar large negative value of $\Delta B_{\Lambda}(10,5)$ was found following a measurement of the excitation spectrum of $^{10}_{\ \Lambda}$Be at JLab~\cite{ref:gogami2} compared to the one obtained for $^{10}_{\ \Lambda}$B by SKS~\cite{ref:hashi}. This observation led the authors of Ref.~\cite{ref:gogami2} to look for a possible systematic bias in the SKS data due to the normalization of all spectra to \BL($^{12}_{\ \Lambda}$C). For this purpose, the differences between the data from the emulsion experiments and those reported by SKS were plotted as a function of $A$. Only the statistical errors listed in Table~\ref{tab:tab1} were considered. Systematic positive values for different $A$ were found, with a weighted average (w.a.) of 540 $\pm$ 50 keV, which could be attributed to an offset on \BL($^{12}_{\ \Lambda}$C) reported by the emulsion experiments~\cite{ref:davis} and taken as reference for all the SKS measurements.
By applying this correction, $\Delta B_{\Lambda}(10,5)$ = +40 $\pm$ 120 keV and $\Delta B_{\Lambda}(12,6)$ = $-$230 $\pm$ 190 keV were found, more compatible with the theoretical expectations.
The need of applying a correction of $\sim$+600 keV to the SKS data was discussed in Ref.~\cite{ref:gal1} from considerations based on a critical analysis of the emulsion data for $^{12}_{\ \Lambda}$C. By assuming that this correction should be the same for all $A$, a revised table of the \BLs values reported by SKS was produced. However, as noticed by the authors themselves, an unexpected large value for $\Delta B_{\Lambda}(16,8)$, $\sim$ $-$700 keV, still persists.
We recall that doubts on the \BL($^{12}_{\ \Lambda}$C) reference value used in SKS data normalization were already raised in Ref.~\cite{ref:cusanno}.

The present paper is structured as follows.
In Sec.~\ref{sec2} we reorganize the data published a few years ago by the FINUDA Collaboration on the spectroscopy of $p$-shell hypernuclei.
In Sec.~\ref{sec3} we compare them with the corresponding ones from the SKS.
We then infer the offset that should be applied to the latter.
Some comments on the emulsion data are given as well.
In Sec.~\ref{sec4} we discuss the results from the present analysis and some final comments are put forward in Sec.~\ref{sec5}.

\section{The \BLs for $p$-shell hypernuclei from FINUDA}
\label{sec2}
Design, features and performance of the FINUDA experiment have been described in general in Ref.~\cite{ref:epja} and specifically in Refs.~\cite{ref:plb622,ref:plb681,ref:npa881,ref:plb698,ref:plb685} as far as the experimental details discussed in this Section are concerned.
The apparatus concept has been developed keeping in mind the specific peculiarity of being a solenoidal magnetic spectrometer for Hypernuclear Physics studies installed on the ($e^{+}$, $e^{-}$) collider DA$\Phi$NE at INFN-LNF.
This feature made possible a data taking strategy completely different from the ones adopted for the SKS and for the double high resolution spectrometers at JLab.
\par
FINUDA exploited the $(K^{-}_{stop}, \pi^{-})$ hypernucleus production reaction on thin nuclear targets. Eight slabs of different materials could be installed closely around the intersection region of the ($e^{+}$, $e^{-}$) colliding beams, tuned at 510 MeV to produce $\phi$ mesons nearly at rest.
The $K^{+}$ and $K^{-}$ pair, emitted back-to-back in the $\phi$ decay (B.R. $\sim$ 49$\%$~\cite{ref:PDG}) with an energy of $\sim$16 MeV, were stopped in the thin nuclear targets.
In this paper we will consider the results obtained out of three $^{12}$C targets (1.7 mm thick, mean density 2.265 g\,cm$^{-3}$), two $^7$Li ones (4.0 mm thick), two $^9$Be ones (2.0 mm thick), a $^{13}$C one (99\% enriched powder, 10.0 mm thick, mean density 0.350 g\,cm$^{-3}$) and a D$_2$O filled one (mylar walled, 3.0 mm thick) as from Ref.~\cite{ref:plb685}.
\par
The apparatus concept permitted to tune different experimental features to measure at best many different physical observables.
For instance, it was the case of the momentum resolution for the charged particles versus the statistical significance of the peaks observed in the spectra.
In this way it was possible to perform specific physical analyses starting from the same raw data sample collected with a minimum bias trigger, by simply applying different sets of selection criteria.
On the contrary, at KEK and at JLab dedicated experimental layouts had to be setup in different runs to pursue different physics goals.

Moreover, FINUDA had the unique advantage of having continuously available during all the runs an unambiguous physical signal for calibration and check  purposes. Since both $K^{+}$ and $K^{-}$ from the $\phi$ decay were stopped in the nuclear targets, many events due to the $K^{+} \rightarrow \mu^{+} + \nu_{\mu}$ decay mode (B.R. $\sim$ 63.51$\%$~\cite{ref:PDG}) were regularly acquired.
The monochromatic $\mu^{+}$, with momentum of 235.535 $\pm$ 0.008 MeV/c~\cite{ref:PDG}, was produced at a rate close to three orders of magnitude larger than the one of the $\pi^{-}$ following the hypernuclei formation. The corresponding physical signal was used for an uninterrupted monitoring of the resolution, of the energy scale with related systematic errors, of the fraction of kaons stopping in the different targets and of the stability of the detector during the long runs, lasting several months.

The spectrometer behavior, as far as the momentum measurement is concerned, was expected to be linear by design and actually the measured magnetic field intensity remained constant in time along the overall data taking period.
Nevertheless, the linearity of the spectrometer was carefully monitored, by checking the reconstructed momentum values of some physical monochromatic signals. In the low momentum region the $\pi^-$ coming from the two-body mesonic weak decay of $^{4}_{\Lambda}$H was considered; the mean value of the Gaussian functions fitting the peaks from the decay at rest of the hyperfragment produced in all targets was used.
In the 200--300 MeV/c momentum interval, the two prongs from the two-body $K^{+} \rightarrow \pi^{+} + \pi^{0}$ and $K^{+} \rightarrow \mu^{+} + \nu_{\mu}$ weak decays were considered for kaons decaying at rest inside all targets.
Finally, in the high momentum region the $\phi \rightarrow e^{+} e^{-}$ decay signal was exploited~\cite{ref:nima570}.
For the latter, the error on the $e^{\pm}$ momentum is very large because it was determined by an on-line procedure, just in order to monitor the collider luminosity.
This essentially means that the number of such events is significantly lower compared to the other three data sets.
In Table~\ref{tab:t0} the expected ($p_\mathrm{exp}$) and measured ($p_\mathrm{meas}$) momenta for these peaks are reported.

\begin{table}[h]
\footnotesize
\center
\caption{Expected and measured momenta characterizing the reactions used to check the linearity of the FINUDA spectrometer.
In the last two columns absolute and relative residues are reported, respectively.}
\label{tab:t0}
\setlength{\extrarowheight}{2.5pt}
\bigskip
\begin{tabular}{|c|c|c|c|c|}
\hline
 reaction & $p_\mathrm{exp}$ & $p_\mathrm{meas}$ & $p_\mathrm{meas} - p_\mathrm{exp}$ & $\frac{p_\mathrm{meas} - p_\mathrm{exp}} {p_\mathrm{exp}}$ \\
 & (MeV/c) & (MeV/c) & (MeV/c) & $\times$ 10$^{-4}$ \\
\hline \hline
 $^{4}_{\Lambda}\mathrm{H} \rightarrow \pi^{-} + \ ^{4}\mathrm{He}$ & 132.9 $\pm$ 0.1~\protect{\cite{ref:esser}} & 132.738 $\pm$ 0.038 & $-$0.16 $\pm$ 0.11 & $-$12 $\pm$ 8 \\
\hline
 $K^{+} \rightarrow \pi^{+} + \pi^{0}$ & 205.138~\protect{\cite{ref:PDG}} & 205.10 $\pm$ 0.01 & $-$0.038 $\pm$ 0.010 & $-$1.9 $\pm$ 0.5 \\
\hline
 $K^{+} \rightarrow \mu^{+} + \nu_{\mu}$ & 235.535~\protect{\cite{ref:PDG}} & 235.410 $\pm$ 0.002 & $-$0.125 $\pm$ 0.002 & $-$5.3 $\pm$ 0.1 \\
\hline
$\phi \rightarrow e^{+} e^{-}$ & 509.730~\protect{\cite{ref:PDG}} & 509.5 $\pm$ 5.0 & $-$0.23 $\pm$ 5.00 & $-$5 $\pm$ 100 \\
\hline
\end{tabular}
\end{table}

From column 3 of Table~\ref{tab:t0} it appears that the residues between $p_\mathrm{meas}$ and $p_\mathrm{exp}$ are all negative, with a w.a. of $-$0.122 $\pm$ 0.002 MeV/c.
Such a value is very close to the residue got in the case of the $K^{+} \rightarrow \mu^{+} + \nu_{\mu}$ decay, which is affected by the lowest error.
A more detailed discussion on the determination of the systematic error for momenta of $\pi^-$ from hypernuclear production will be addressed at the end of this Section.
Fig.~\ref{fig:fig0} shows the curve fitted to the four points and it reports the linear fit parameters.

\begin{figure}
\begin{center}
\includegraphics[width=0.75\textwidth]{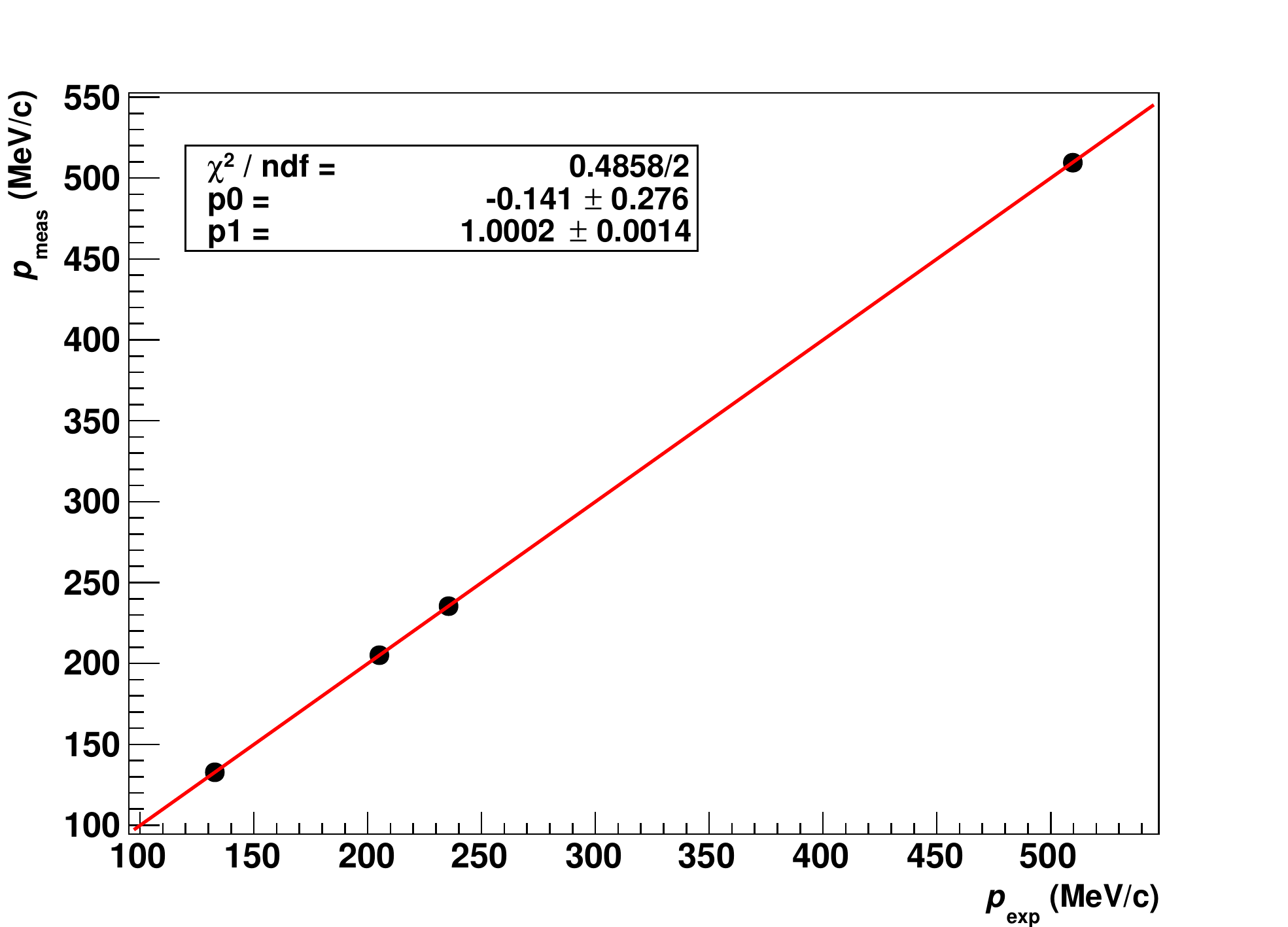}
\caption {Linearity curve of the FINUDA spectrometer. $p_\mathrm{exp}$ indicates the expected momentum, $p_\mathrm{meas}$ the measured one for the four calibration reactions listed in Table~\ref{tab:t0}.}
\label{fig:fig0}
\end{center}
\end{figure}
%


In column 5 of Table~\ref{tab:tab1} we have listed pairs of \BLs values from FINUDA for $^{7}_{\Lambda}$Li, $^{9}_{\Lambda}$Be and $^{12}_{\ \Lambda}$C, reason for which we will briefly comment in the following. The first paper on hypernuclear spectroscopy by FINUDA~\cite{ref:plb622} reported the excitation spectrum of $^{12}_{\ \Lambda}$C, with the aim of confirming the existence of peaks due to excited states between the two main signals corresponding to the \LLs in $s$-shell (ground state) and in $p$-shell.
For this purpose the main effort was put on the achievement of the best energy resolution by requiring only high quality tracks with a good $\chi^{2}$ value. These tracks correspond to two categories, namely $\pi^{-}$'s emitted in the forward hemisphere with respect to the stopping $K^{-}$ direction for the hypernuclear production and $\mu^{+}$'s emitted in the forward hemisphere with respect to the stopping $K^{+}$ direction for calibration.
Actually, in this case particles pass through a minimum amount of material before entering the tracker. A resolution of 1.29 MeV FWHM was obtained, the best up to now achieved in meson-induced hypernuclear production experiments. The fit of the experimental spectrum to states with energies corresponding to those given in Ref.~\cite{ref:hotchi} was not satisfactory. A better result could be obtained by searching for a new set of ground and excited states energies. However, at that time the energy calibration was not yet optimized since the spectra from only two out of the three $^{12}$C targets used in the run could be added up. The energy scale for the third one was displaced by 0.5 MeV, which we may assume as systematic error for the results of Ref.~\cite{ref:plb622}.

The goal of a subsequent analysis~\cite{ref:plb681} was to measure the energy spectra of $\pi^{-}$ from the mesonic decay of some $p$-shell hypernuclei. They were measured in coincidence with the $\pi^{-}$ from the formation reaction which identified the ground state. In this case the main requirement was to have the largest number of events of interest. To this end, the selection criteria on the formation $\pi^-$ tracks were less stringent, with a worsening of the energy resolution, 2.31 MeV FWHM for excitation spectra of $^{7}_{\Lambda}$Li, $^{9}_{\Lambda}$Be and $^{12}_{\ \Lambda}$C. However, an increase of the number of events in the ground state peaks by a factor 6 was achieved~\cite{ref:npa804}. The mean values of the Gaussian curves best fitting the peaks corresponding to the ground state of the hypernuclei are reported in column 5 of Table~\ref{tab:tab1}, with the statistical errors. The systematic errors were not evaluated in Ref.~\cite{ref:plb681}, since not relevant to that analysis. An evaluation of the maximum systematic uncertainty affecting the determination of the $K_{\mu2}$ signal was given in Ref.~\cite{ref:npa881}.
It amounts to 0.1 MeV as reported in column 5 of Table~\ref{tab:tab1}.
In the case of $^{16}_{\ \Lambda}$O the fit to the ground state has a lower quality, due to the reduced statistics and to the poorer resolution. Its \BLs was fixed at 12.42 MeV, as given in Ref.~\cite{ref:hashi}.

In these inclusive $\pi^{-}$ spectra, it is possible to recognize peaks corresponding to the formation of hyperfragments: $^{11}_{\ \Lambda}$B from $^{12}$C targets and $^{15}_{\ \Lambda}$N from the $^{16}$O target.
These peaks have been, indeed, used to tag the mesonic decay of the hyperfragments.
However, it is not possible to determine the \BLs values of their ground states since the hyperfragment production peaks only provide the masses of the excited resonant states of the parent hypernuclei.
Nevertheless, from the measurement of the energy of the $\pi^{-}$ from the mesonic decay (which always occurs from the ground state) the \BLs value can be obtained.
Unfortunately the achievable precision is limited due to the reduced statistics of the sample and to the poorer resolution on the low energy decay $\pi^{-}$'s, typically $\sim$4 MeV FWHM.
The evaluation of \BLs by means of the spectroscopy of $\pi^-$ from the mesonic decay was not done in Ref.~\cite{ref:plb681}, focused at that time on the determination of observables which characterize the weak decay process.
Nevertheless, by applying this method to $^{7}_{\Lambda}$Li and to $^{9}_{\Lambda}$Be we find \BLs = 5.70 $\pm$ 0.25 MeV and \BLs = 6.88 $\pm$ 0.76 MeV, respectively.
Such values are in good agreement with the more precise values obtained by the spectroscopy of $\pi^-$ from the hypernucleus formation reaction (see Table~\ref{tab:tab1}).
These less precise values will not be used in the present discussion, while the values obtained for $^{11}_{\ \Lambda}$B and $^{15}_{\ \Lambda}$N will be taken into account since they represent the only determinations obtained with calibrated magnetic spectrometers.
They are reported in column 5 of Table~\ref{tab:tab1} with statistical and systematic errors.
In particular, the latter has been inferred from the maximum variation of the hyperfragment mass obtained when varying initial values in the peak fitting procedure, within three times the size of the error got in the first iteration step.
It is worth to remind that the determination of the produced hypernucleus mass starting from the mesonic decay requires the knowledge of the daughter excitation spectrum which has been assumed from theoretical calculations~\cite{ref:galmes}.

A further contribution from FINUDA to the spectroscopy of $p$-shell hypernuclei can be found in Ref.~\cite{ref:plb698}. By taking advantage of the substantial progress achieved in the effectiveness of the corrections to be applied to the raw data, excitation spectra for $^{7}_{\Lambda}$Li, $^{9}_{\Lambda}$Be, $^{13}_{\ \Lambda}$C and $^{16}_{\ \Lambda}$O were obtained by requiring high quality tracks, like in Ref.~\cite{ref:plb622}, but including in the analysis also backward emitted $\pi^{-}$, which passed through larger thickness of materials. The statistics was nearly doubled, the resolution was 1.76 MeV FWHM and the systematic error on the energy was 0.3 MeV. A total error of 0.4 MeV on the peak centers, as reported in Ref.~\cite{ref:plb698}, is quoted in column 5 of Table~\ref{tab:tab1}. It appears that there is a substantial agreement between the two sets of values published by FINUDA. However, we cannot use both of them to evaluate a w.a. since they were deduced from the same data sample.
For $^{7}_{\Lambda}$Li, $^{9}_{\Lambda}$Be and $^{12}_{\ \Lambda}$C only the results from Ref.~\cite{ref:plb681}, affected by smaller errors, will be then taken into account.

It is worth to make clear that the size of the systematic error affecting the different \BLs determinations by FINUDA depends on the strictness of the criteria adopted to select tracks. In other words, the progressive application of more and more stringent constraints unavoidably introduces additional biases on the sample of tracks taken into account.
Indeed, this is the reason for the increase of the systematic error on \BLs from Ref.~\cite{ref:plb681} to Ref.~\cite{ref:plb698} and finally to Ref.~\cite{ref:plb622}.
\par
Fig.~\ref{fig:fig0b} shows the distribution of the $\mu^{+}$ momentum obtained when the selection criteria of Ref.~\cite{ref:plb681} are applied.
For simplicity, it refers to one of the eight targets. The green line histogram (Pmag) is the spectrum of the raw momentum obtained from the spectrometer before applying corrections for energy loss in materials, while the black line histogram (Pmod) is the final spectrum. The (blue) vertical line indicates the $\mu^+$ nominal momentum value. For each track a correction of the raw reconstructed momentum was applied by taking into account the corresponding energy loss mean value evaluated along its own measured trajectory, by the Bethe-Bloch formula without the density effect term. The method was validated by means of a Monte Carlo simulation which described very carefully the geometry of the apparatus and the material budget. The effectiveness of this correction is evident from the momentum distributions of Fig.~\ref{fig:fig0b}.
\begin{figure}[h]
\begin{center}
\includegraphics[width=0.75\textwidth]{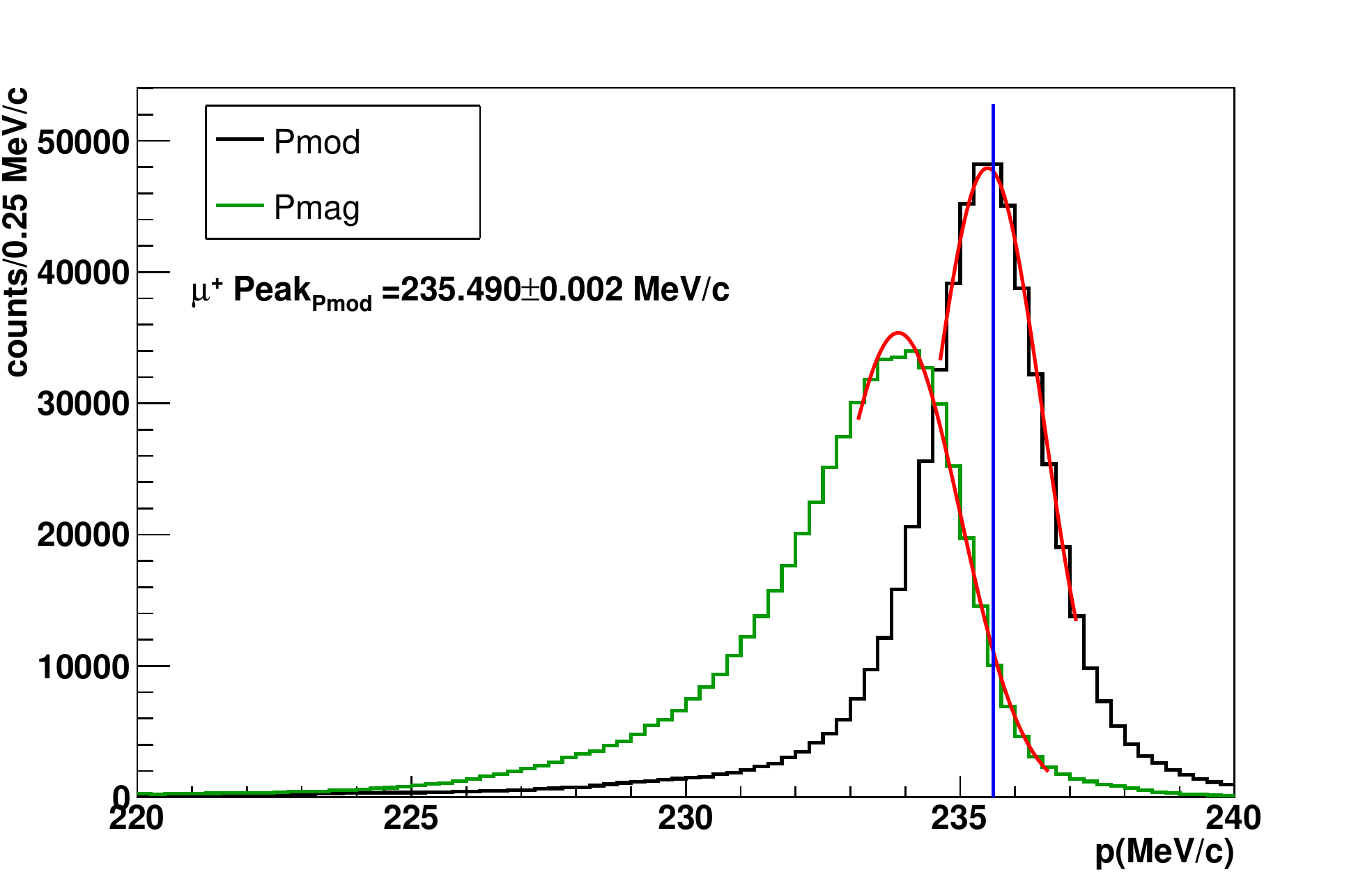}
\caption{Spectrum of the $\mu^{+}$ momentum from the $K^{+} \rightarrow \mu^{+} + \nu_{\mu}$ reaction at rest reconstructed by FINUDA, before (green histogram) and after (black histogram) the correction for energy loss in targets and detector materials. The (blue) vertical line indicates the $\mu^+$ nominal momentum value. The (red) curves are Gaussian function fits to the peaks.}
\label{fig:fig0b}
\end{center}
\end{figure}
The spectra correspond to the total sample of the 2006-2007 data taking period. Both spectra feature an asymmetric shape due to two main contributions: the radiative $K^{+} \rightarrow \mu^{+} + \nu_{\mu} + \gamma$ decay channel, which amounts to $\sim 10 \%$ of the non-radiative one~\cite{ref:PDG}, and the high energy part of the energy loss distribution. The asymmetry is then quite sizeable for the uncorrected spectrum. On the other hand, it is also present, even though to a lesser extent, in the corrected spectrum since it cannot be completely eliminated by the mean energy loss correction which was applied track by track. Therefore, a Gaussian function fit over an asymmetric fitting range was performed to obtain the most probable momentum. The fitting function is shown on both spectra in Fig.~\ref{fig:fig0b}, while its mean value is only reported for the corrected peak (235.490 $\pm$ 0.002 MeV/c). We got similar results for all targets, leading to the w.a. value listed in Table~\ref{tab:t0}. By considering each single target, the maximum difference w.r.t. the nominal value was 0.12 MeV/c, which was considered as the systematic error on the momentum. Under the assumption of the perfect linearity of the FINUDA spectrometer, expected by design and experimentally verified (see Fig.~\ref{fig:fig0}), such a value corresponds to 0.1 MeV systematic error on the energy of the pion from hypernuclear production as reported in column 5 of Table~\ref{tab:tab1} in the case of the results from Ref.~\cite{ref:plb681}.
Anyway, it has to be underlined that these numbers represent pessimistic estimates because they have been extracted from the cumulative distributions collected during the whole data taking period.
Hence, they could include possible time shifts in the energy calibration of the spectrometer and in this respect their small value establishes a robust upper limit for such an effect.

The data collected by FINUDA allowed also to search for neutron-rich hypernuclei produced in two-body reactions such as:
\begin{equation}
              K^{-}_{stop} + \, ^{A}Z \rightarrow \,^{A}_{\Lambda}(Z - 2) + \pi^{+}.
\label{nrich}
\end{equation}
Their existence would be signalled by the presence of narrow peaks in the measured $\pi^{+}$ spectra, with momenta in the 250--260 MeV/c region. The observation is difficult since the capture rate for reaction (\ref{nrich}) is $\leq 10^{-2}$ w.r.t. the one for the production of $^{A}_{\Lambda}Z$ hypernuclei and $\pi^{+}$ peaks are blurred into the spectra of $\pi^{+}$ coming from other processes.
From a first analysis only upper limits on the production of $^{6}_{\Lambda}$H, $^{7}_{\Lambda}$H and $^{12}_{\ \Lambda}$Be could be deduced~\cite{ref:plb640}.
In a following analysis the $\pi^{+}$ spectra from production reactions on $^{6}$Li targets were observed in coincidence with the $\pi^{-}$ from the $^{6}_{\Lambda}$H $\rightarrow\ ^{6}$He + $\pi^{-}$ weak decay.
Three unambiguous events were found in a larger data sample, leading to a determination of \BL($^{6}_{\Lambda}\mathrm{H})= 4.00\pm1.1$ MeV~\cite{ref:prl108}.
A full account of all the analysis procedures and simulations can be found in Ref.~\cite{ref:npa881}.
In particular, the error quoted in Ref.~\cite{ref:prl108} is total and the systematic contribution is 0.12 MeV.

\section{Comparison of the FINUDA results on \BLs with previous measurements}
\label{sec3}
The series of \BLs values from FINUDA for $_\Lambda^7$Li, $_\Lambda^9$Be, $_{\ \Lambda}^{12}$C, $_{\ \Lambda}^{13}$C and $_{\ \Lambda}^{16}$O (column 5 of Table~\ref{tab:tab1}) can be directly compared with the analogous results obtained by the SKS Collaboration (column 3 of Table~\ref{tab:tab1}).
As anticipated in the Introduction, the energy scale of the SKS was calibrated by the physical signal corresponding to the ground state of $_{\ \Lambda}^{12}$C, with \BLs = 10.80 MeV, taken from the emulsion data~\cite{ref:davis}.
Several authors~\cite{ref:gal1,ref:gogami2,ref:cusanno} pointed out that the above procedure of calibration applied to all the \BLs from SKS led to difficulties in the interpretation of the data and some of them suggested alternative approaches.
Since the \BL's from FINUDA are measured with a spectrometer with an accurate absolute energy scale, a comparison between the two sets of data (columns 3 and 5) would provide the normalization factor for the energy scale of SKS.
For this purpose, the differences between the \BL's from FINUDA and SKS were evaluated (+0.63 $\pm$ 0.18, +0.31 $\pm$ 0.16, +0.77 $\pm$ 0.11, $-$0.38 $\pm$ 0.40, +0.98 $\pm$ 0.40 MeV) for the previously listed hypernuclei.
They are shown in Fig.~\ref{fig:fig1}.

\begin{figure}[h]
\begin{center}
\includegraphics[width=\textwidth]{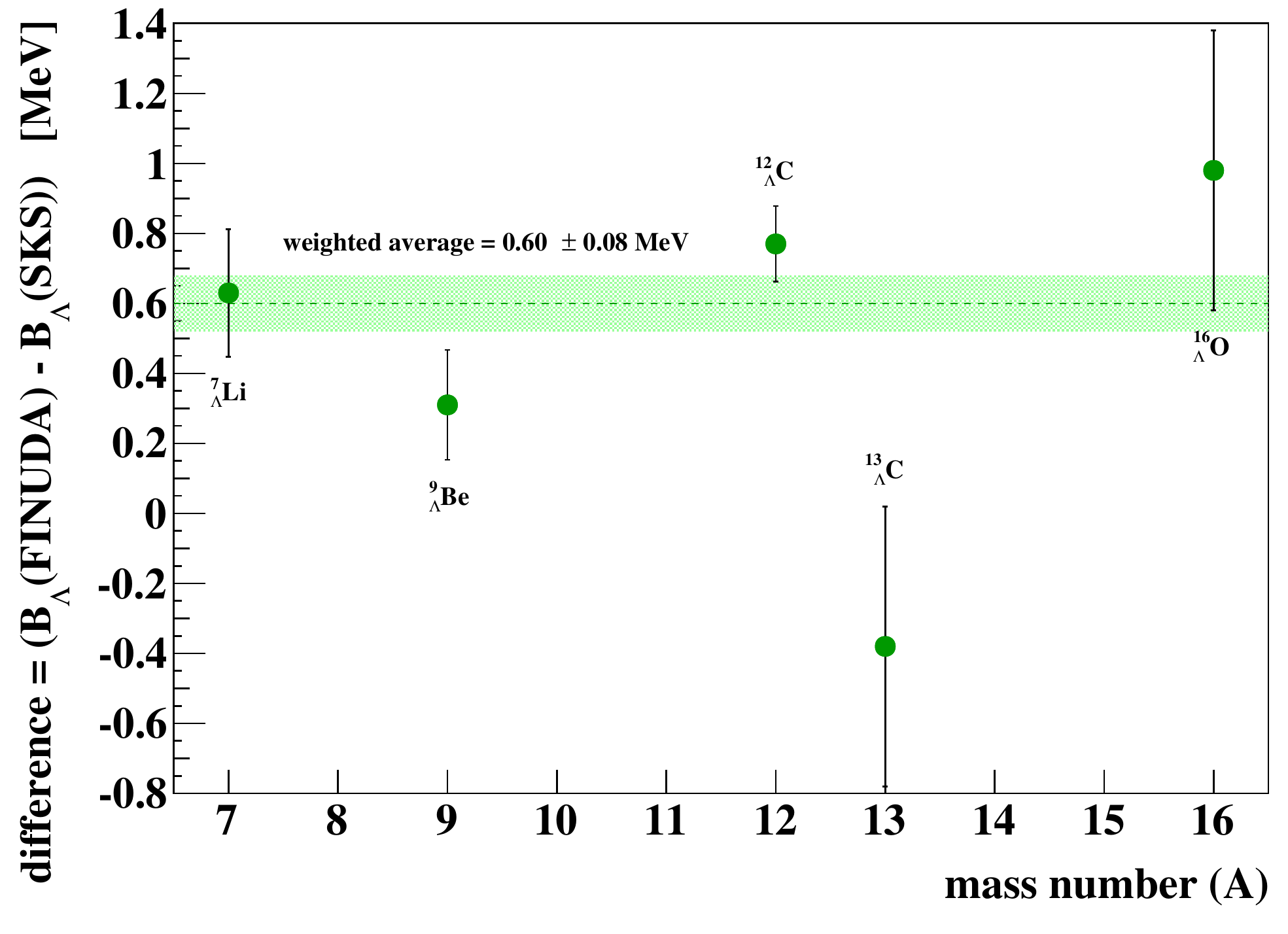}
\caption{Binding energy differences between the FINUDA data (column 5 of Table~\ref{tab:tab1}) and the analogous SKS data (column 3 of Table~\ref{tab:tab1}) for $^{7}_{\Lambda}$Li, $^{9}_{\Lambda}$Be, $^{12}_{\ \Lambda}$C, $^{13}_{\ \Lambda}$C and $^{16}_{\ \Lambda}$O. The errors taken into account are total for the FINUDA results and statistical only in the SKS case. The weighted average of +600 $\pm$ 80 keV is represented by the (green) dashed line.}
\label{fig:fig1}
\end{center}
\end{figure}

The errors taken into account are total for the FINUDA data and statistical only for the SKS data.
It should be noted that the systematic errors on the \BLs from FINUDA are not the same for all hypernuclei, as they are for the SKS.
The w.a. of these differences is +600 $\pm$ 80 keV. 
It is very close to the normalization value suggested by Ref.~\cite{ref:gal1} and it agrees well within the errors with the offset evaluated in Ref.~\cite{ref:gogami2}.
The latter was obtained with a procedure similar to the one described before, in which the w.a. of the differences between the \BLs reported by the emulsion experiments (column 2 of Table~\ref{tab:tab1}, Ref.~\cite{ref:davis}) and by SKS (column 3 of Table~\ref{tab:tab1}, Ref.~\cite{ref:hashi}) were evaluated.
However, only the statistical errors were considered for both sets of data.
Hence, the final number should be considered as the difference between the presumably equal systematic error affecting all the \BLs from SKS and a possible common systematic error for all the \BLs from emulsion measurements.
A plausible source of systematic error on the emulsion data could be that $M$(\LL), which enters directly into the definition of \BLs (see Eq.~(\ref{BL})), changed from 1115.57 $\pm$ 0.03 MeV, measured and adopted in Ref.~\cite{ref:juric}, to 1115.683 $\pm$ 0.006 MeV~\cite{ref:PDG}, used in recent spectrometric experiments.
Naively, one may guess that a correction of +113 keV should be applied to the emulsion data when compared to recent spectrometric measurements.
However, there is a quite subtle interplay in the emulsion technique between the measurements of $M$(\LL)$c^2$ and $M(^{A}_{\Lambda}Z)c^2$ which appear in Eq.~(\ref{BL})~\cite{ref:wilquet}.
Actually, $M(^{A}_{\Lambda}Z)c^2$ was determined exclusively from $\pi^-$ mesonic decay and $M$(\LL)$c^2$ from the $p \pi^-$ decay in the same emulsion stack.
Since the above masses appear with opposite signs in Eq.~(\ref{BL}), systematic errors possibly occurring in the range-energy relationship for $\pi^-$ (emulsion density first of all) were thus partially compensated by selecting events with comparable $\pi^-$ energies from \LLs and $_{\Lambda}^A\!Z$ mesonic decays.
The systematic error of $\pm$40 keV mentioned in Ref.~\cite{ref:davis} probably accounts for this effect.
\par
We tried to determine a possible systematic error in all the emulsion data by using the same approach described before in the case of the SKS results.
The differences between the values from spectrometers with an absolute energy scale (FINUDA and electroproduction, columns 5 and 6 of Table~\ref{tab:tab1}) and the corresponding ones from emulsions were calculated.
The error was evaluated by using the total error for the spectrometric measurements and the statistical one for the emulsion data.
The w.a. of the differences was found to be +57 $\pm$ 44 keV with a $\chi^2$/d.o.f. = 22.07/8 = 2.76 when the compilation of Ref.~\cite{ref:juric} is taken into account and +79 $\pm$ 43 keV with a $\chi^2$/d.o.f. = 34.56/9 = 3.84 when the compilation of Ref.~\cite{ref:davis} is considered.
Both values of the obtained reduced $\chi^2$ show that the simplified assumption of a common systematic error on the emulsion data is not valid.

\section{CSB in $p$-shell hypernuclei}
\label{sec4}
\begin{figure}
\begin{center}
\includegraphics[width=\textwidth]{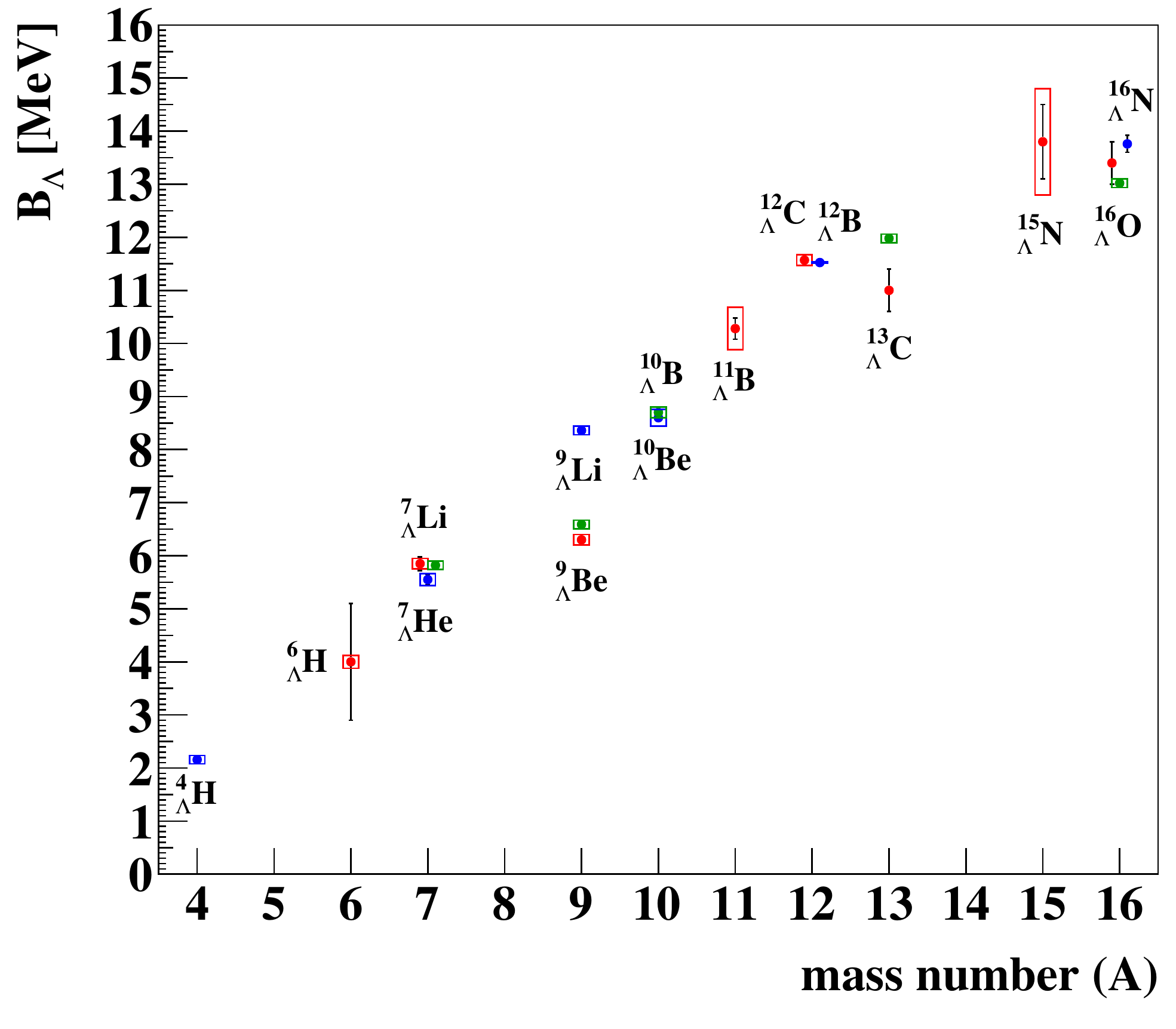}
\caption{Summary plot of the \BLs for $A \leq 16$ \LL-hypernuclei from columns 4, 5 and 6 of Table~\ref{tab:tab1}.
The green markers indicate the revised SKS results; the red and blue ones correspond to the values provided by the FINUDA and the JLab and MaMi electroproduction experiments, respectively.
Vertical bars are the statistical uncertainty while boxes represent the systematic error.}
\label{fig:fig3}
\end{center}
\end{figure}
In this Section we will compare the \BLs in the case of isomultiplets as obtained from calibrated counter experiments (FINUDA and electroproduction).
In the discussion we will refer to the renormalized SKS data.
They are obtained by adding 600 keV to the original values and by considering $\pm$80 keV as systematic error, following the analysis described at the beginning of Sec.~\ref{sec3}.
They are reported in column 4 of Table~\ref{tab:tab1} (SKS rev.).
Fig.~\ref{fig:fig3} summarizes the recent results provided by counter experiments about the \BL's of $A \leq 16$ hypernuclei.
\par
The archetype of the CSB effect in the T = 1/2, $A$ = 4 hypernuclei dates back to the beginning of Hypernuclear Physics.
A quite large $\Delta B_\Lambda$(4,2) = +350 $\pm$ 50 keV was in fact reported from the analysis of the emulsion data~\cite{ref:davis}.
The result was confirmed by a recent measurement of $\Delta B_\Lambda$(4,2) carried out for the first time with the novel technique of the high resolution pion decay spectroscopy at MaMi~\cite{ref:esser} (see Table~\ref{tab:tab1}).
The new determination reaffirmed a substantial ground state (J = 0) splitting $\Delta B_\Lambda$(4,2) = +270 $\pm$ 95 keV due to CSB, which is consistent with the emulsion value cited above.
A comparison with a very recent theoretical calculation of the CSB in the $A$ = 4 hypernuclei can be found in Refs.~\cite{ref:gazda,ref:gazda2}.
The same conclusion on the existence and on the magnitude of the CSB effect in the T = 1/2, $A$ = 4 hypernuclear isodoublet was drawn by the E13 Collaboration at J-PARC thanks to the measurement of the $\gamma$-rays from the 1$^+ \rightarrow$ 0$^+$ M1 transition in $_\Lambda^4$He~\cite{ref:yamamoto}.
\par
The recent identification of $_\Lambda^7$He and the measurement of its \BL~\cite{ref:nakamura,ref:gogami1} stimulated a strong interest for possible CSB effects in the lowest-$A$ $p$-shell hypernuclear system.
$_\Lambda^7$He, $_\Lambda^7$Li$^*$ and $_\Lambda^7$Be belong to the same $A$ = 7, T = 1 isotriplet.
The corresponding $B_\Lambda$'s given by different experiments are reported in Table~\ref{tab:tab1} and shown by Fig.~\ref{fig:fig5}, where the bars represent statistical errors, the red hatched areas systematic errors.
\begin{figure}[t]
\begin{center}
\includegraphics[width=0.9\textwidth]{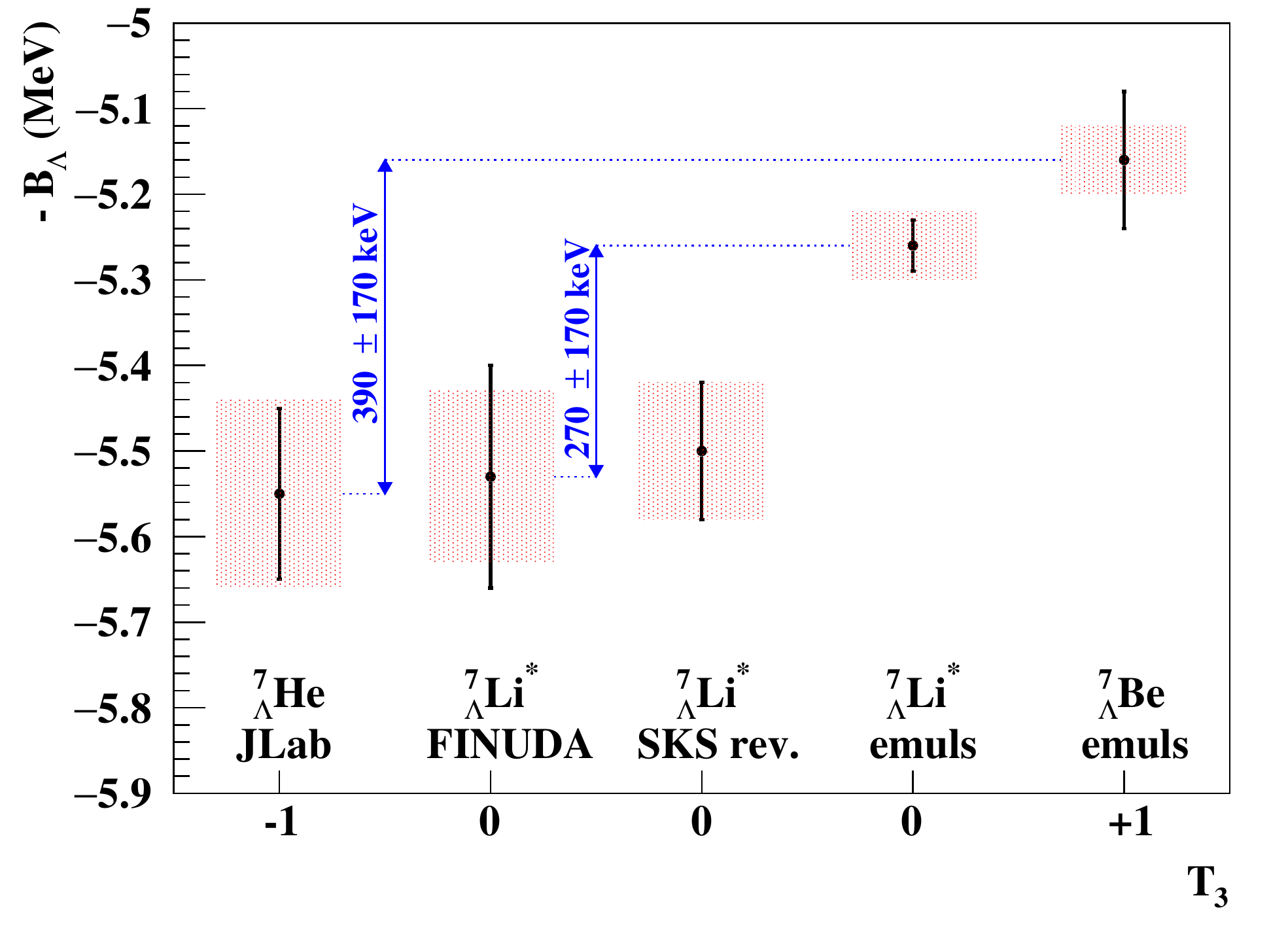}
\caption{Measured \LLs binding energies for the hyperisobars of the $A$ = 7, T = 1 isotriplet ($^{7}_{\Lambda}$He, $^{7}_{\Lambda}$Li*, $^{7}_{\Lambda}$Be). The vertical bars represent statistical errors, the red hatched areas systematic errors.}
\label{fig:fig5}
\end{center}
\end{figure}
We observe that the \BLs difference between the two extreme members T$_3$ = +1,$-$1 of the isotriplet is $-$390 $\pm$ 170 keV.
It could be even much lower since $-$270 $\pm$ 170 keV are due to the still existing difference between the \BL($_\Lambda^7$Li$^*$) determinations provided by counter and emulsion experiments.
A further remark is that, according to Ref.~\cite{ref:hiyama}, a difference of $-$150 keV between \BL($_\Lambda^7$He) and \BL($_\Lambda^7$Be) is expected to originate from the Coulomb force, which causes a different shrink of the nuclear cores.
With the caution arising from the still large errors we conclude that no substantial contributions from the CSB effect seem to be present in the $A$ = 7, T = 1 hypernuclear isotriplet.
\par
Table~\ref{tab:tab3} gives a summary of the $\Delta B_\Lambda$($A$,$Z$) measured for the isomultiplets of the $p$-shell hypernuclei listed in column 1.
Column 2 reports the values of $\Delta B_\Lambda$($A$,$Z$) which may be deduced from the data of Table~\ref{tab:tab1}.
In column 3 the sources of the experimental information are specified with pertinent References reported in column 4.
For sake of completeness, we remind that among $p$-shell hypernuclei there is also the isomultiplet with $A$ = 9.
The difference between the \BLs values for $_\Lambda^9$B and $_\Lambda^9$Li, measured in emulsion experiments~\cite{ref:davis}, amounts to $-$210 $\pm$ 220 keV.
However, we don't include this value in Table ~\ref{tab:tab3} because in this case $\Delta Z$ = 2, contrary to what we assumed in Eq.~(\ref{DBL}).
\begin{table}[h]
\center
\caption{$\Delta B_\Lambda$($A$,$Z$) (column 2) measured for isomultiplet pairs of observed $p$-shell hypernuclei (column 1). Among the sources of the experimental data listed in column 3, SKS indicates the values revised from the original results of SKS normalized to the emulsion data, as done in Ref.~\protect{\cite{ref:gogami2}} (see Sec.~\protect{\ref{sec3}}). Finally, the Reference paper from which the $\Delta B_\Lambda$($A$,$Z$) was taken is indicated in column 4.}
\label{tab:tab3}
\setlength{\extrarowheight}{3.5pt}
\bigskip
\begin{tabular}{|c|c|c|c|}
\hline
multiplet pair & $\Delta B_\Lambda(A,Z)\  \mathrm{(keV)} $           & experimental sources            & Reference \\
\hline  \hline
$_\Lambda^7$Be         $-$ $_\Lambda^7$Li$^*$     & $-$100 $\pm$  90 & emuls. $-$ emuls.               & \protect{\cite{ref:davis,ref:tamura}} \\                                                                              \hline
$_\Lambda^7$Li$^*$     $-$ $_\Lambda^7$He         &  $-$20 $\pm$ 230 & FINUDA $-$ ($e$,$e^\prime K^+$) & [t.w.] \\                                                                              \hline
$_\Lambda^8$Be         $-$ $_\Lambda^8$Li         &    +40 $\pm$  60 & emuls. $-$ emuls.               & \protect{\cite{ref:davis}} \\                                                                              \hline
$_{\ \Lambda}^{10}$B   $-$ $_{\ \Lambda}^{10}$Be  & $-$220 $\pm$ 250 & emuls. $-$ emuls.               & \protect{\cite{ref:davis}} \\
                                                  &    +40 $\pm$ 120 & SKS $-$ ($e$,$e^\prime$,$K^+$)  & \protect{\cite{ref:gogami2}} \\
\hline
$_{\ \Lambda}^{12}$C   $-$ $_{\ \Lambda}^{12}$B   & $-$570 $\pm$ 190 & emuls. $-$ emuls.               & \protect{\cite{ref:davis}} \\
                                                  & $-$230 $\pm$ 190 & SKS $-$ ($e$,$e^\prime$,$K^+$)  & \protect{\cite{ref:gogami2}} \\
                                                  &    +50 $\pm$ 110 & FINUDA $-$ ($e$,$e^\prime K^+$) & [t.w.] \\
\hline
$_{\ \Lambda}^{16}$O   $-$ $_{\ \Lambda}^{16}$N   & $-$360 $\pm$ 430 & FINUDA $-$ ($e$,$e^\prime K^+$) & [t.w.] \\                                                                              \hline
\end{tabular}
\end{table}

We add a few comments.
The large value reported for $\Delta B_\Lambda$(12,6) in Ref.~\cite{ref:davis} is due to the problem of $B_\Lambda$($_{\ \Lambda}^{12}$C) in the emulsion data previously discussed.
As far as $_{\ \Lambda}^{16}$O is concerned, we remark that the authors of Ref.~\cite{ref:cusanno} noted that the value of \BLs by SKS, 12.42 $\pm$ 0.05 MeV, may be underestimated when compared with the updated values 13.3 $\pm$ 0.4 MeV and 13.4 $\pm$ 0.4 MeV from previous experiments with the ($K^-$,$\pi^-$) reaction both at rest and in flight.
The argument was furthermore mentioned in Ref.~\cite{ref:plb698}, in discussing the spectrum of excitation of $_{\ \Lambda}^{16}$O.
The use of the normalized SKS value, given in column 4 of Table~\ref{tab:tab1}, attenuates the discrepancy but still leads to a sizeable value of $\Delta B_\Lambda$(16,8) suggesting an unexpectedly large CSB effect of the order of +700 keV, very hard to be explained theoretically~\cite{ref:gal1}.
For this reason in Table~\ref{tab:tab3} we only report the $\Delta B_\Lambda$(16,8) obtained by comparing the FINUDA and the JLab measurements.
\par
From Table~\ref{tab:tab3} it follows that all data indicate a small, if any, $\Delta B_\Lambda$($A$,$Z$) for $p$-shell hypernuclei.
CSB in the $\Lambda N$ interaction seems to be smaller than in $s$-shell hypernuclei, supporting a recent prediction by Gal~\cite{ref:gal2}.
The present evaluation of the CSB effect for the $A$ = 7, 12, 16 hypernuclear multiplets is the first one based on the published results by recent experiments with magnetic spectrometers featuring an absolute energy scale calibration.
Data are not corrected for the Coulomb force contribution.

\section{Summary and outlook}
\label{sec5}
We have reorganized the recent data from FINUDA on the spectroscopy of $p$-shell hypernuclei in order to compare them with the ones of the neutron-rich isobars from JLab.
For both sets of data the energy scale exploits an absolute calibration and searches for CSB effects in $p$-shell hypernuclei are thus reliable.
The pattern of the energy levels for all members of the T = 1, $A$ = 7 hypernuclear isotriplet ($^7_\Lambda$He, $^7_\Lambda$Li$^*$, $^7_\Lambda$Be) was examined by combining the results from JLab, FINUDA and emulsion measurements.
There is a discrepancy between the values of $^7_\Lambda$Li$^*$ obtained by the FINUDA and the emulsion experiments.
$\Delta B_\Lambda$($A$,$Z$) for $A$ = 7, 12, 16 were deduced by using only data provided by magnetic spectrometers with an absolute scale calibration.
In all of the three cases $\Delta B_\Lambda$ is consistent with zero within a 100--200 keV error, being much smaller than in the $A$ = 4 case as predicted by recent theoretical evaluations.
\par
In recent years, a strong theoretical effort has been put in predicting several interesting effects related to the CSB of $p$-shell and higher-$A$ hypernuclei.
It should then be useful to start a similar effort on the experimental side.
For the neutron-rich hyperisobars the complexes of spectrometers and the associated technologies developed for the study of the electroproduction reaction are well suited for the purpose.
On the contrary, for the hypernuclei obtained by using the ($K^-$,$\pi^-$) and ($\pi^+$,$K^+$) reactions a new experimental approach should be started.
J-PARC is able to provide excellent beams of $K^-$ and of $\pi^+$, but a new generation of spectrometers must be developed.
Actually, they should feature a resolution of some hundreds of keV and, first of all, they must rely on a methodology capable of providing an absolute energy calibration to the best of 100 keV.
\par
With regards to nuclear targets to be used, the first choice should be $^4$He, urgently needed to confirm $B_\Lambda$($_\Lambda^4$He), based today only on emulsion data.
A very important target would be $^7$Li, for a further confirmation of the presently available spectrometric data on $B_\Lambda$($_\Lambda^7$Li).
The same argument holds for a target of $^{16}$O.
\par
The experimental study of the T = 1, $A$ = 7 isotriplet should be completed by a new value of $B_\Lambda$($_\Lambda^7$Be) with counter measurements.
However, this task looks very hard with the present accelerator machines and detection technologies.
\par
The new generation of spectrometers should be capable of measuring with improved precision the \BLs of $p$-shell hypernuclei produced by the two-body meson-induced reactions out of $^AZ$ nuclear targets.
In this way it should be possible to address, and hopefully to settle, the issue of the present discrepancies between old emulsion data and recent spectrometric measurements (see Table~\ref{tab:tab1}).
\par
Obviously, investigations would and should be extended to hypernuclei belonging to other shells as well.
\par
As far as the $s$-shell is concerned, \BL($_\Lambda^3$H) could not be measured by exploiting ($K^-$,$\pi^-$) or ($\pi^+$,$K^+$) reactions due to the radiation safety requirements imposed for the handling of $^3$H radioactive targets.
In our opinion it is important to determine \BL($_\Lambda^3$H) by means of high precision counter experiments.
$_\Lambda^3$H is the weakest few-body bound system of hadrons with strangeness.
Also in this case, its \BL = 0.13 $\pm$ 0.05 $\pm$ 0.04 MeV was measured only by emulsion experiments.
\par
In a recent paper~\cite{ref:npa954} the possibility of exploiting the $^3$He($\pi^-$,$K^0$)$_\Lambda^3$H reaction is studied.
The high flux $\pi^-$ beam available at J-PARC combined with the SKS spectrometer, used to detect the $\pi^+$ from the asymmetric $K^0 \rightarrow \pi^+\pi^-$ decay, seems to offer a realistic possibility concerning the rate of production of $_\Lambda^3$H, as needed for the precise measurement of the lifetime of $_\Lambda^3$H.
Unfortunately, this is not true for the measurement of \BL($_\Lambda^3$H) due to the limited missing mass resolution ($\sim$ 3 MeV).
\par
A realistic experimental approach to the precision measurement of \BL($_\Lambda^3$H) seems to rely on the electroproduction reaction on a $^3$He target.
Production of $_\Lambda^3$H with the ($e$,$e^\prime K^+$) reaction on $^3$He was already observed in a first generation experiment at JLab featuring a missing mass resolution of 4 MeV~\cite{ref:dohrmann}.
Preliminary evaluations of the event rate which is expected with the high resolution spectrometers in operation at JLab may be found in Ref.~\cite{ref:hyp15}.
Another experimental approach could be the high resolution pion decay spectroscopy of $_\Lambda^3$H obtained as hyperfragment in electroproduction reactions.

\bigskip
We are indebted to Prof. G. Wilquet for the very useful and clarifying correspondence about the methodology of the emulsions technique. Prof. A. Gal is acknowledged for qualified comments and remarks. Prof. D.H. Davis provided helpful information.





\end{document}